\pdfoutput=1
\documentclass[12pt,a4paper]{article}

\usepackage{lineno}  
\usepackage{graphicx}  
\usepackage{xspace} 
\usepackage{color}
\usepackage{colortbl}
\usepackage{amsmath} 
\usepackage{ifthen} 
\usepackage{multirow}
\graphicspath{{./figs/}} 

\newboolean{pdflatex}
\setboolean{pdflatex}{true} 
\newboolean{articletitles}
\setboolean{articletitles}{true} 

\newboolean{uprightparticles}
\setboolean{uprightparticles}{false} 

\usepackage{amssymb}
\usepackage{amsfonts}
\usepackage{upgreek} 

\usepackage{hyperref}    

\usepackage[all]{hypcap} 

\graphicspath{{figures/}}
\def\lhcb {\mbox{LHCb}\xspace}
\def\ux85 {\mbox{UX85}\xspace}

\ifthenelse{\boolean{uprightparticles}}%
{

 \def\Ppsi        {\ensuremath{\uppsi}\xspace}

 \def\PDelta      {\ensuremath{\Delta}\xspace}                 
 \def\PXi      {\ensuremath{\Xi}\xspace}                 
 \def\PLambda      {\ensuremath{\Lambda}\xspace}                 
 \def\PSigma      {\ensuremath{\Sigma}\xspace}                 
 \def\POmega      {\ensuremath{\Omega}\xspace}                 
 \def\PUpsilon      {\ensuremath{\Upsilon}\xspace}                 
 
 \def\PB      {\ensuremath{\mathrm{B}}\xspace}                 
                  
 \def\PD      {\ensuremath{\mathrm{D}}\xspace}

 \def\PJ      {\ensuremath{\mathrm{J}}\xspace}                 
 \def\PK      {\ensuremath{\mathrm{K}}\xspace}

 \def\Pi      {\ensuremath{\mathrm{i}}\xspace}

}
{

 \def\Ppsi        {\ensuremath{\psi}\xspace}                 
                  
 \mathchardef\PDelta="7101
 \mathchardef\PXi="7104
 \mathchardef\PLambda="7103
 \mathchardef\PSigma="7106
 \mathchardef\POmega="710A
 \mathchardef\PUpsilon="7107
                  
 \def\PB      {\ensuremath{B}\xspace}                 
                  
 \def\PD      {\ensuremath{D}\xspace}

 \def\PJ      {\ensuremath{J}\xspace}                 
 \def\PK      {\ensuremath{K}\xspace}

 \def\Pi      {\ensuremath{i}\xspace}

}

\def\kaon  {\ensuremath{\PK}\xspace}
  \def\Kbar  {\kern 0.2em\overline{\kern -0.2em \PK}{}\xspace}

\def\Kz    {\ensuremath{\kaon^0}\xspace}
\def\Kzb   {\ensuremath{\Kbar^0}\xspace}
\def\KzKzb {\ensuremath{\Kz \kern -0.16em \Kzb}\xspace}
\def\Kp    {\ensuremath{\kaon^+}\xspace}
\def\Km    {\ensuremath{\kaon^-}\xspace}

\def\KpKm  {\ensuremath{\Kp \kern -0.16em \Km}\xspace}

  \def\Dbar    {\kern 0.2em\overline{\kern -0.2em \PD}{}\xspace}
\def\D       {\ensuremath{\PD}\xspace}

\def\Dz      {\ensuremath{\D^0}\xspace}
\def\Dzb     {\ensuremath{\Dbar^0}\xspace}
\def\DzDzb   {\ensuremath{\Dz {\kern -0.16em \Dzb}}\xspace}
\def\Dp      {\ensuremath{\D^+}\xspace}
\def\Dm      {\ensuremath{\D^-}\xspace}

\def\DpDm    {\ensuremath{\Dp {\kern -0.16em \Dm}}\xspace}

\def\B       {\ensuremath{\PB}\xspace}
  \def\Bbar    {\kern 0.18em\overline{\kern -0.18em \PB}{}\xspace}

\def\Bz      {\ensuremath{\B^0}\xspace}

\def\Bs      {\ensuremath{\B^0_\squark}\xspace}
\def\Bsb     {\ensuremath{\Bbar^0_\squark}\xspace}

\def\jpsi     {\ensuremath{{\PJ\mskip -3mu/\mskip -2mu\Ppsi\mskip 2mu}}\xspace}

  \def\Y#1S{\ensuremath{\PUpsilon{(#1S)}}\xspace}

\def\Lbar {\ensuremath{\kern 0.1em\overline{\kern -0.1em\PLambda}}\xspace}

\newcommand{\decay}[2]{\ensuremath{#1\!\to #2}\xspace}         
\def\ra                 {\ensuremath{\rightarrow}\xspace}
\def\to                 {\ensuremath{\rightarrow}\xspace}

\def\CP                {\ensuremath{C\!P}\xspace}

\newcommand{\sinphis}{\ensuremath{\sin\!\phis}\xspace}

\def\AT#1     {\ensuremath{A_{\mathrm{T}}^{#1}}\xspace}           

\def\C#1      {\ensuremath{\mathcal{C}_{#1}}\xspace}                       
\def\Cp#1     {\ensuremath{\mathcal{C}_{#1}^{'}}\xspace}                    
\def\Ceff#1   {\ensuremath{\mathcal{C}_{#1}^{\mathrm{(eff)}}}\xspace}        
\def\Cpeff#1  {\ensuremath{\mathcal{C}_{#1}^{'\mathrm{(eff)}}}\xspace}       
\def\Ope#1    {\ensuremath{\mathcal{O}_{#1}}\xspace}                       
\def\Opep#1   {\ensuremath{\mathcal{O}_{#1}^{'}}\xspace}                    

\newcommand{\tev}{\ensuremath{\mathrm{\,Te\kern -0.1em V}}\xspace}
\newcommand{\gev}{\ensuremath{\mathrm{\,Ge\kern -0.1em V}}\xspace}
\newcommand{\mev}{\ensuremath{\mathrm{\,Me\kern -0.1em V}}\xspace}
\newcommand{\kev}{\ensuremath{\mathrm{\,ke\kern -0.1em V}}\xspace}
\newcommand{\ev}{\ensuremath{\mathrm{\,e\kern -0.1em V}}\xspace}
\newcommand{\gevc}{\ensuremath{{\mathrm{\,Ge\kern -0.1em V\!/}c}}\xspace}
\newcommand{\mevc}{\ensuremath{{\mathrm{\,Me\kern -0.1em V\!/}c}}\xspace}
\newcommand{\gevcc}{\ensuremath{{\mathrm{\,Ge\kern -0.1em V\!/}c^2}}\xspace}
\newcommand{\gevgevcccc}{\ensuremath{{\mathrm{\,Ge\kern -0.1em V^2\!/}c^4}}\xspace}
\newcommand{\mevcc}{\ensuremath{{\mathrm{\,Me\kern -0.1em V\!/}c^2}}\xspace}

\def\gsim{{~\raise.15em\hbox{$>$}\kern-.85em
          \lower.35em\hbox{$\sim$}~}\xspace}
\def\lsim{{~\raise.15em\hbox{$<$}\kern-.85em
          \lower.35em\hbox{$\sim$}~}\xspace}

\newcommand{\Real}{\ensuremath{\mathcal{R}e}\xspace}
\newcommand{\Imag}{\ensuremath{\mathcal{I}m}\xspace}

\def\tell1  {TELL1\xspace}
\def\ukl1   {UKL1\xspace}

\def \t {\theta}

\def \ra {\rangle}
\def \Bs {B_s^{0}}

\def \Bq {B_q^{0}}
\def \Bqb {\overline{B}{}_q^{0}}
\def \Bsb {\overline{B}{}_s^{0}}
\def \G {\Gamma}

\def \ch {\cosh \frac{\Delta \G t}{2}}
\def \cs {\cos(\Delta m t)}
\def \sh {\sinh \frac{\Delta \G t}{2}}
\def \sn {\sin(\Delta m t)}

\def \Ab {\overline{A}}
\def \angmu {\theta_{\ell}}
\def \angpi {\theta_{h}}
\def\tone {\theta_{\ell}}
\def\ttwo {\theta_{h}}
\def \CH {{\cal H}}
\def \CHb {{\cal \overline{H}}}
\def \App {A_{\perp}}
\def \Apl {A_{\parallel}}
\def \trt {\theta_{\rm tr}}
\def \trphi {\phi_{\rm tr}}
\def \trpsi {\psi_{\rm tr}}
\def \dv {{\rm d}}
\usepackage{cite}
\usepackage{mciteplus}

\usepackage{epsfig,accents}
\newcommand*{\fancybar}{\scalebox{.4}{(}\raisebox{-1.7pt}{--}\scalebox{.4}{)}}
\newcommand*{\brabar}[1]{\accentset{\fancybar}{#1}}

\begin{document}
\renewcommand{\thefootnote}{\fnsymbol{footnote}}
\setcounter{footnote}{1}



\begin{titlepage}
\pagenumbering{roman}
\belowpdfbookmark{Title page}{title}

\pagenumbering{roman}
\hspace*{-5mm}\begin{tabular*}{16cm}{lc@{\extracolsep{\fill}}r}
 & & December 27, 2012 \\ 
 & & \\
\hline
\end{tabular*}

\vspace*{2.5cm}

{\bf\boldmath\huge
\begin{center}
Time-dependent Dalitz-plot formalism for $B_q^0\rightarrow \jpsi h^+ h^-$
\end{center}
}

\vspace*{2.0cm}
\begin{center}
Liming Zhang and  Sheldon Stone
\bigskip\\
{\it\footnotesize
Physics Department
Syracuse University, Syracuse, NY, USA 13244-1130\\

}
\end{center}


\begin{abstract}
  \noindent
A formalism for measuring time-dependent \CP violation in $\B^0_{(s)}\rightarrow \jpsi h^+h^-$ decays with  $\jpsi\to \mu^+\mu^-$  is developed for the general case where there can be many $h^+h^-$ final states of different angular momentum present. Here $h$ refers to any spinless meson.
The decay amplitude is derived using similar considerations as those in a Dalitz like analysis of three-body spinless mesons taking into account the fact that the \jpsi is spin-1, and the various interferences allowed between different final states. Implementation of this procedure can, in principle, lead to the use of a larger number of final states for \CP violation studies.

\end{abstract}

\vspace*{2.5cm}

\newpage

\end{titlepage}

\renewcommand{\thefootnote}{\arabic{footnote}}
\setcounter{footnote}{0}

\pagestyle{empty}  



\setcounter{page}{2}
\mbox{~}




\pagestyle{plain} 
\setcounter{page}{1}
\pagenumbering{arabic}


%
\section{Introduction}
Measurement of \CP violation in the $\Bz$ and $\Bs$ systems is important for testing the Standard Model, as new particles can appear in mixing diagrams. Previous measurements have been made in many modes \cite{PDG}. To measure the phase in $\Bs$ decays the final states $\Bs\to \jpsi K^+K^-$ for $K^+K^-$ masses close to that of the $\phi$ meson has been used \cite{LHCb-CONF-2012-002,LHCb:2011aa,CDF:2011af,*Abazov:2011ry,*:2012fu}, as well as
 $\Bs\to \jpsi \pi^+\pi^-$ \cite{LHCb:2012ad,*LHCb:2011ab}. In the latter case the final state is \CP odd \cite{LHCb:2012ae} over most of the $\pi^+\pi^-$ mass range, while in the case of $K^+K^-$ the final state even in the mass region near the $\phi$ meson has both \CP odd and even components, that can be resolved using time-dependent angular analysis \cite{Dighe:1995pd,*Dighe:1998vk}. In this paper we present a formalism that allows the entire $K^+K^-$ mass region to be used in \CP violation measurements regardless of the final state angular momentum. This formalism can also be applied to \Bz decays, e.g. $\Bz\to \jpsi \pi^+\pi^-$.

The basic concept here is to couple a three-body Dalitz like analysis
\cite{Dalitz:1953cp} to the $\jpsi h^+ h^-$ final state, where the $\jpsi\to\mu^+\mu^-$ and concurrently measure the time-dependent \CP violation by splitting the final state into odd and even \CP components.

\section{Time-dependent decay rates}
The time evolution of the $B^0_q$-$\Bqb$ system is described by the
Schr\"{o}dinger equation
\begin{equation}
i\frac{\partial}{\partial
t}\left(\begin{array}{c}{|B^0_q}(t)\rangle\\{|\Bqb}(t)\rangle\end{array}\right)=
\left(\mbox{\boldmath $\rm M$}-\frac{i}{2}\mbox{\boldmath $\rm
\Gamma$}\right)
\left(\begin{array}{c}{|B^0_q}(t)\rangle\\{|\Bqb}(t)\rangle\end{array}\right),
\end{equation}
where the \mbox{\boldmath $\rm M$} and \mbox{\boldmath $\rm \Gamma$}
matrices are Hermitian, and $CPT$ invariance implies that
$M_{11}=M_{22}$ and $\Gamma_{11}=\Gamma_{22}$.
The off-diagonal elements, $M_{12}$ and $\Gamma_{12}$, of these matrices describe the off-shell (dispersive)
 and on-shell (absorptive) contributions
to $B^0_q$-$\Bqb$ mixing, respectively.

The mass eigenstates $|B_H\rangle$ and $|B_L\rangle$ of the effective Hamiltonian matrix are given by
\begin{eqnarray}
|B_L\ra &=& p|B_q^0\ra+q|\Bqb\ra, \nonumber\\
|B_H\ra &=& p|B_q^0\ra-q|\Bqb\ra,
\end{eqnarray}
with $|p|^2+|q|^2=1$. The decay amplitudes for $\B_q^0$ and $\Bqb$ into a self-charge-conjugated final state $f$, where for this paper $f=\jpsi h^+ h^-$, are defined as
\begin{equation}
 A_f \equiv \langle f|S|B_q^0\ra, \quad \quad \quad \Ab_f \equiv \langle f|S|\Bqb\ra.
\end{equation}
With the additional definitions
\begin{equation}
A \equiv A_f, \quad {\rm and} \quad \Ab \equiv \frac{q}{p}\Ab_f,
\end{equation}
the time dependent decay rates can be written as \cite{Nierste:2009wg,*Bigi:2000yz}
\begin{eqnarray}\label{Eq-t}
\Gamma(t) =\quad\quad
  {\cal N} e^{-\G t}\left\{\frac{|A|^2+|\Ab|^2}{2}\ch  + \frac{|A|^2-|\Ab|^2}{2}\cs\right.\quad\quad\quad\quad\quad\quad\quad\quad\nonumber\\
- \left.\Real(A^*\Ab)\sh  -  \Imag(A^*\Ab)\sn\right\},\\
\overline{\Gamma}(t) =
 \left|\frac{p}{q}\right|^2{\cal N}  e^{-\G t}\left\{\frac{|A|^2+|\Ab|^2}{2}\ch  - \frac{|A|^2-|\Ab|^2}{2}\cs\right.\quad\quad\quad\quad\quad\quad\quad\quad\nonumber\\
 - \left.\Real(A^*\Ab)\sh  +  \Imag(A^*\Ab)\sn\right\},\label{Eqbar-t}
\end{eqnarray}
where $\cal N$ is a normalization constant, $\Delta m= m_H-m_L$, $\Delta \Gamma = \Gamma_L-\Gamma_H$, and $\Gamma = (\Gamma_L+\Gamma_H)/2$.

\section{Angular dependent formulas}
\subsection{Definition of helicity angles}

\begin{figure}[ht]
  \centering
  \includegraphics[width=.9\textwidth]{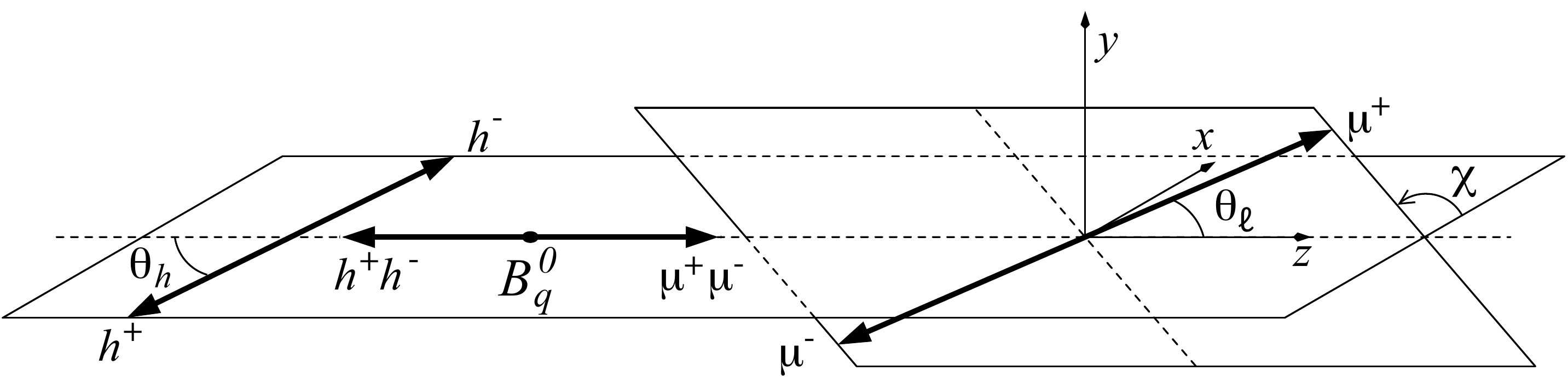} 
  \caption{Definition of helicity angles. For details see text.}
  \label{fig:helicityAngles}
\end{figure}

We express the angular dependence of the decay in terms of ``helicity" angles defined as
(i) $\angmu$, the angle between the $\mu^+$ direction in the $\jpsi$ rest frame with respect to the $\jpsi$ direction in the $\Bq$ rest frame;  (ii) $\angpi$ the angle between the $h^+$ direction in the $h^+h^-$ rest frame with respect to the $h^+h^-$ direction in the $\Bq$ rest frame,
and (iii) $\chi$ the angle between the $\jpsi$ and $h^+h^-$ decay planes in the $\Bq$ rest frame. These angles are shown pictorially in Fig.~\ref{fig:helicityAngles}. (These definitions are the same for $\Bq$ and $\Bqb$, namely, using $\mu^+$ and $h^+$ to define the angles for both $\Bq$ and $\Bqb$ decays.)
\subsection{Time-independent part of the rate for $B^0_q$ decays}
For the decays of $\B_q^0\to\jpsi h^+h^-$ with $\jpsi\to \mu^+\mu^-$ the decay rate is found by summing over the unobserved lepton polarizations. The time-independent part of the rate is\footnote{In $d_{-\lambda,0}^J(\angpi)$, $-\lambda$ is used instead of $\lambda$ in order to be consistent with the convention used in~\cite{LHCb:2011aa}.}
\begin{equation}\label{Eq7}
|A_f(m_{hh},\angpi,\angmu,\chi)|^2
=\sum_{\alpha=\pm1}\left|\sum^{|\lambda|\leq J}_{\lambda,J}\sqrt{\frac{2J+1}{4\pi}}H_{\lambda}^J(m_{hh}) e^{i\lambda\chi}d_{\lambda,\alpha}^1(\angmu)d_{-\lambda,0}^J(\angpi)\right|^2,
\end{equation}
where $\lambda=0,\pm1$ is the $\jpsi$ helicity, $\alpha=\pm1$ is the helicity difference between the two muons, $J$ is the spin of the $h^+h^-$ intermediate state, and $H_{\lambda}^J(m_{hh})$ is a helicity amplitude depending on $m_{hh}$ that can be expressed using a formalism similar to that in a Dalitz-plot analyses.
We define the term which contains the sum over spin-$J$ as
\begin{equation}\label{H}
{\cal H}_{\lambda}(m_{hh},\angpi) = \sum_J \sqrt{\frac{2J+1}{4\pi}}H_{\lambda}^J(m_{hh}) d_{-\lambda,0}^J(\angpi).
\end{equation}
Then Eq.~(\ref{Eq7}) becomes
\begin{align}
&\hspace{-12mm}|A_f(m_{hh},\angpi,\angmu,\chi)|^2
=\sum_{\alpha=\pm1}\left|\sum_{\lambda} e^{i\lambda\chi}d_{\lambda,\alpha}^1(\angmu) {\cal H}_{\lambda}(m_{hh},\angpi) \right|^2 \nonumber\\
&\hspace{5mm}= \sum_{\alpha=\pm1}\left[\left(\sum_{\lambda^\prime} e^{i\lambda^\prime\chi} d_{\lambda^\prime,\alpha}^1(\angmu) {\cal H}_{\lambda^\prime}(m_{hh},\angpi) \right)^*\left(\sum_{\lambda} e^{i\lambda\chi} d_{\lambda,\alpha}^1(\angmu) {\cal H}_{\lambda}(m_{hh},\angpi) \right)\right]\nonumber\\
&\hspace{5mm}=\sum_{\lambda^\prime,\lambda}\left(\sum_{\alpha=\pm1} d_{\lambda^\prime,\alpha}^1(\angmu)d_{\lambda,\alpha}^1(\angmu) \right){\cal H}_{\lambda^\prime}^{ *}(m_{hh},\angpi){\cal H}_{\lambda}(m_{hh},\angpi) \;e^{i(\lambda-\lambda^\prime)\chi}.
\end{align}
Defining
\begin{equation}\label{Eq:Theta}
\Theta_{\lambda^\prime\lambda}(\angmu)\equiv\displaystyle\sum_{\alpha=\pm1}d_{\lambda^\prime,\alpha}^1(\angmu)d_{\lambda,\alpha}^1(\angmu),
\end{equation}
results in
\begin{equation}\label{9}
|A_f(m_{hh},\angpi,\angmu,\chi)|^2=\sum_{\lambda^\prime,\lambda}{\cal H}_{\lambda}(m_{hh},\angpi){\cal H}_{\lambda^\prime}^{ *}(m_{hh},\angpi) \;e^{i(\lambda-\lambda^\prime)\chi}\Theta_{\lambda^\prime\lambda}(\angmu).
\end{equation}

Table \ref{Theta} lists the functions $\Theta_{\lambda^\prime\lambda}(\angmu)$. They are invariant under the interchange of $\lambda$ and $\lambda^\prime$, i.e.  $\Theta_{\lambda^\prime\lambda}(\angmu)=\Theta_{\lambda\lambda^\prime}(\angmu)$, and transform with respect to a change of the sign of both $\lambda$ and $\lambda^\prime$ as $\Theta_{\lambda\lambda^\prime}(\angmu)=(-1)^{\lambda-\lambda^\prime}\Theta_{-\lambda^\prime-\lambda}(\angmu)$.  Inserting the explicit functional forms  in Eq.~(\ref{9}) allows us to express the amplitude as
\begin{align}\label{Eq:A}
|A_f(m_{hh},\angpi,\angmu,\chi)|^2=&|\CH_0(m_{hh},\angpi)|^2\sin^2\angmu +\frac{1}{2}\left(|\CH_+(m_{hh},\angpi)|^2+|\CH_-(m_{hh},\angpi)|^2\right)\nonumber\\
&\times (1+\cos^2\angmu)+\Real\left[\CH_+(m_{hh},\angpi)\CH_-^*(m_{hh},\angpi)e^{2i\chi}\right]\sin^2\angmu \nonumber\\
&+\sqrt{2}\Real\left[\left(\CH_0(m_{hh},\angpi)\CH_+^*(m_{hh},\angpi)-\CH^*_0(m_{hh},\angpi)\CH_-(m_{hh},\angpi)\right)e^{-i\chi}\right]\nonumber\\
&\times \sin\angmu\cos\angmu ,
\end{align}
where we denote $\CH_{\lambda}$ by 0, +, and $-$, rather than 0, $+1$ and $-1$.
\begin{table}[hbtp]
\centering
\caption{Functional forms of $\Theta_{\lambda^\prime\lambda}(\theta)$ defined in Eq.~(\ref{Eq:Theta}) for different values of $\lambda$ and $\lambda^\prime$.}\label{Theta}
\def\arraystretch{1.2}
\begin{tabular}{|c|c|r|}\hline
$\lambda$ & $\lambda^\prime$ & $\Theta_{\lambda^\prime\lambda}(\theta)$\\\hline\hline
~~0 & ~~0 & $\sin^2\t$ \\
~~0 &  ~~1 & $\frac{1}{\sqrt{2}}\sin\t\cos\t$\\
~~0 & $-$1& $-\frac{1}{\sqrt{2}}\sin\t\cos\t$\\
~~1 & ~~0 & $\frac{1}{\sqrt{2}}\sin\t\cos\t$\\
~~1 & ~~1 & $\frac{1}{2}(1+\cos^2\t)$\\
~~1 & $-$1 & $\frac{1}{2}\sin^2\t$\\
$-$1& ~~0 & $-\frac{1}{\sqrt{2}}\sin\t\cos\t$\\
$-$1 & ~~1 & $\frac{1}{2}\sin^2\t$\\
$-$1 & $-$1 & $\frac{1}{2}(1+\cos^2\t)$\\\hline
\end{tabular}
\end{table}

\subsection{Time-independent part of the rate for $\overline{B}^0_q$ decays}
For $\overline{B}^0_q$ decays, the expression for $|\Ab_f(m_{hh},\angpi,\angmu,\chi)|^2$, results from replacing $\CH_{\lambda}(m_{hh},\angpi)$ in Eq.~(\ref{Eq:A}) by $\CHb_{\lambda}(m_{hh},\angpi)$, which contains the helicity amplitudes for $\Bqb$ decays. $\CH_{\lambda}(m_{hh},\angpi)$ and $\CHb_{\lambda}(m_{hh},\angpi)$ are related by transversity \CP eigenstates \cite{Dunietz:1990cj}, that are discussed in Section~\ref{final}. Using these we find
\begin{align}\label{Eq:Ab}
|\Ab_f(m_{hh},\angpi,\angmu,\chi)|^2=&|\CHb_0(m_{hh},\angpi)|^2\sin^2\angmu+\frac{1}{2}\left(|\CHb_+(m_{hh},\angpi)|^2+|\CHb_-(m_{hh},\angpi)|^2\right) \nonumber\\
&\times (1+\cos^2\angmu)+\Real\left[\CHb_+(m_{hh},\angpi)\CHb_-^*(m_{hh},\angpi)e^{2i\chi}\right]\sin^2\angmu \nonumber\\
&+\sqrt{2}\Real\left[\left(\CHb_0(m_{hh},\angpi)\CHb_+^*(m_{hh},\angpi)-\CHb^*_0(m_{hh},\angpi)\CHb_-(m_{hh},\angpi)\right)e^{-i\chi}\right] \nonumber\\
&\times \sin\angmu\cos\angmu~.
\end{align}

\subsection{The interference term}
Next we calculate the complex term $A_f^*(m_{hh},\angpi,\angmu,\chi)\Ab_f(m_{hh},\angpi,\angmu,\chi)$. We have
\begin{align}
&\hspace{-12mm}A_f^*(m_{hh},\angpi,\angmu,\chi)\Ab_f(m_{hh},\angpi,\angmu,\chi) =\nonumber\\
&\hspace{10mm}\sum_{\alpha=\pm1}\left[\left(\sum_{\lambda^\prime} e^{i\lambda^\prime\chi} d_{\lambda^\prime,\alpha}^1(\angmu) \CH_{\lambda^\prime}(m_{hh},\angpi) \right)^*\left(\sum_{\lambda} e^{i\lambda\chi} d_{\lambda,\alpha}^1(\angmu) \CHb_{\lambda}(m_{hh},\angpi) \right)\right]\nonumber\\
&\hspace{6mm}=\sum_{\lambda^\prime,\lambda}\CHb_{\lambda}(m_{hh},\angpi){\cal H}_{\lambda^\prime}^{ *}(m_{hh},\angpi) \;e^{i(\lambda-\lambda^\prime)\chi}\Theta_{\lambda^\prime\lambda}(\angmu).
\end{align}
Replacing the explicit terms leads to
\begin{align}\label{Eq:AAb}
&\hspace{-10mm}A_f^*(m_{hh},\angpi,\angmu,\chi)\Ab_f(m_{hh},\angpi,\angmu,\chi)
= \CHb_0(m_{hh},\angpi)\CH_0^*(m_{hh},\angpi)\sin^2\angmu \nonumber\\
&+\frac{1}{2}\left(\CHb_+(m_{hh},\angpi)\CH_+^*(m_{hh},\angpi)+\CHb_-(m_{hh},\angpi)\CH_-^*(m_{hh},\angpi)\right)(1+\cos^2\angmu)\nonumber\\
&+\frac{1}{2}\left(\CHb_+(m_{hh},\angpi)\CH_-^*(m_{hh},\angpi)e^{2i\chi}+\CHb_-(m_{hh},\angpi)\CH_+^*(m_{hh},\angpi)e^{-2i\chi}\right)\sin^2\angmu\nonumber\\
&+\frac{1}{\sqrt{2}}\left(\CHb_0(m_{hh},\angpi)\CH_+^*(m_{hh},\angpi)e^{-i\chi}-\CHb_0(m_{hh},\angpi)\CH_-^*(m_{hh},\angpi)e^{i\chi}\right. \nonumber\\
&\left.+\CHb_+(m_{hh},\angpi)\CH_0^*(m_{hh},\angpi)e^{i\chi}-\CHb_-(m_{hh},\angpi)\CH_0^*(m_{hh},\angpi)e^{-i\chi}\right)\sin\angmu\cos\angmu.
\end{align}

\section{Time-dependent Dalitz-plot formalism}\label{final}
Here we discuss the general formalism which includes S, P, D or higher waves of the $h^+h^-$ intermediate states. 

Apart from the proper decay-time $t$, the decay of $\B_q^0\to \jpsi h^+h^-$, $\jpsi\to\mu^+\mu^-$ can be described by four variables, we choose to use $m_{hh}$ and three helicity angles ($\angmu,\angpi,\chi$), where ($m_{hh}, \cos\angpi$) space is used instead of the usual variables in a  Dalitz-plot analysis: $m^2_{hh}, m^2_{\jpsi h^+}$; the advantage is the former has an rectangle phase space which is easier for calculating the normalization.

Assuming $|p/q|=1$, the differential decay rates in  Eqs.~(\ref{Eq-t}) and~(\ref{Eqbar-t}) can be written  in terms of the five variables $t,m_{hh},\angmu,\angpi,\chi$ as
\begin{eqnarray}
\frac{\dv^5 \Gamma}{\dv t\dv m_{hh} \dv \cos\angmu \dv \cos\angpi \dv \chi } \propto
 e^{-\G t}\left\{\frac{|A_f|^2+|\Ab_f|^2}{2}\ch  + \frac{|A_f|^2-|\Ab_f|^2}{2}\cs\right.\nonumber\\
- \left.\Real\left(\frac{q}{p}A_f^*\Ab_f\right)\sh  -  \Imag\left(\frac{q}{p}A_f^*\Ab_f\right)\sn\right\},\label{eq:pdf1}\\
\frac{\dv^5 \overline{\Gamma}}{\dv t\dv m_{hh} \dv \cos\angmu \dv \cos\angpi \dv \chi } \propto
 e^{-\G t}\left\{\frac{|A_f|^2+|\Ab_f|^2}{2}\ch  - \frac{|A_f|^2-|\Ab_f|^2}{2}\cs\right.\nonumber\\
- \left.\Real\left(\frac{q}{p}A_f^*\Ab_f\right)\sh  + \Imag\left(\frac{q}{p}A_f^*\Ab_f\right)\sn\right\}\label{eq:pdf2}.
\end{eqnarray}


The functions $|A_f|^2$, $|\Ab_f|^2$ and $A_f^*\Ab_f$ are defined in Eqs.~(\ref{Eq:A}),~(\ref{Eq:Ab}) and~(\ref{Eq:AAb}) respectively. We now substitute in Eq.~(\ref{H}) explicit variables for $H_{\lambda}^J(m_{hh})$ in terms of our chosen Dalitz-plot variables $m_{hh}$ and $\angpi$, resulting in
\begin{equation}
\label{eq:heart}
{\cal \brabar{H}}_{\lambda}(m_{hh},\angpi) = \sum_R  \brabar{\bf h}_\lambda^R  \sqrt{2J_R+1} \sqrt{P_R P_B}  F_B^{(L_B)} F_R^{(L_R)} A_R(m_{hh})\left(\frac{P_B}{m_B}\right)^{L_B}\left(\frac{P_R}{m_{hh}}\right)^{L_R} d_{-\lambda,0}^{J_R}(\angpi),
\end{equation}
where the function $A_R(m_{hh})$ describes the mass squared shape of the resonance $R$, that in most cases is a Breit-Wigner function, $P_B$ is the \jpsi   momentum in the $\overline{B}^0_q$ rest frame, $P_R$ is the momentum of either of the two hadrons in the dihadron rest frame, $m_{B}$ is the $\overline{B}^0_q$ mass, $L_B$ is the orbital angular momentum between the $J/\psi$ and $h^+h^-$ system, and $L_R$ the orbital angular momentum in the $h^+h^-$ decay, and thus is the same as the spin of the $h^+h^-$ resonance. $F_B^{(L_B)}$ and $F_R^{(L_R)}$ are the Blatt-Weisskopf barrier factors for $\overline{B}^0_q$ and $R$ resonance respectively~\cite{LHCb:2012ae}.

The factor $\sqrt{P_RP_B}$ results from converting the phase space of  the natural Dalitz-plot variables $m^2_{hh}$ and  $m^2_{\jpsi h^+}$ to that of $m_{hh}$ and $\cos\angpi$. ${\cal \brabar{H}}_{\lambda}$ is summed over all $h^+h^-$ intermediate states ($R$) with different spins, denoted as $J_R$. The function defined in Eq.~(\ref{eq:heart}) is based on previous Dalitz plot analyses \cite{Mizuk:2008me, LHCb:2012ae}, but here all allowed values of $L_B$ and $L_R$ are included.

In order to use \CP relations, it is convenient to replace the helicity complex coefficients $\brabar{\textbf{h}}_\lambda^R$ by the transversity complex coefficients $\brabar{\textbf{a}}_{i}^R$ using their relations
\begin{eqnarray}
\brabar{\textbf{a}}_{0}^R &=& \brabar{\textbf{h}}_0^R,\nonumber\\
\brabar{\textbf{a}}_\parallel^R &=& \frac{1}{\sqrt{2}}(\brabar{\textbf{h}}_+^R+\brabar{\textbf{h}}_-^R),\nonumber\\
\brabar{\textbf{a}}_\perp^R &=& \frac{1}{\sqrt{2}}(\brabar{\textbf{h}}_+^R-\brabar{\textbf{h}}_-^R).
\end{eqnarray}
Here $\brabar{\textbf{a}}_{0}^R$ corresponds to longitudinal polarization of the $\jpsi$ meson, and the other two coefficients correspond to polarizations of the $\jpsi$ meson and $h^+h^-$ system transverse to the decay axis: $\brabar{\textbf{a}}_\parallel^R$ for parallel polarization of the $\jpsi$ and $h^+h^-$ and $\brabar{\textbf{a}}_\perp^R$ for their perpendicular polarization.

In the SM, if we assume that only one diagram contributes to the decay and there is no direct \CP violation, then the CKM weak phase only appears as $\frac{q}{p}=e^{-i\phi_s}$ for the $\Bs$ decays and $\frac{q}{p}=e^{-2i\beta}$ for the $\Bz$ decays. The  $\brabar{\textbf{a}}_i$ amplitudes only contain strong phases, so $\bar{\textbf{a}}^R_i=\eta^R_i\textbf{a}^R_i$, where $\eta^R_i$ is \CP eigenvalue of the $i$th  transversity component for the intermediate state $R$. (Here $i= 0,~\parallel,~\perp$.)
Note that for the $h^+h^-$ system both $C$ and $P$ are given by
$(-1)^{L_R}$, so the  \CP of  the $h^+h^-$ system is always even. The total CP of the final state is $(-1)^{L_B}$, since the \CP of the \jpsi is also even.
The final state \CP parities for S, P,  and D-waves are shown in Table~\ref{CPPart}.

\begin{table}
\centering
\caption{\CP parity for different spin resonances. Note that spin-0 only has one transversity component-$0$.}\label{CPPart}
\begin{tabular}{c|ccc}\hline
Spin& $\eta_0$& $\eta_\parallel$& $\eta_\perp$\\\hline
0 & $-1$ & ~-- &~--\\
1 & ~~1 & ~~1 & $-1$\\
2 & $-1$ & $-1$ & ~~1\\ \hline
\end{tabular}
\end{table}


Direct \CP violation can also be considered, i.e. $\bar{\textbf{a}}^R_i \neq \eta^R_i\textbf{a}^R_i$. The complex coefficients can be parameterized as 
\begin{equation}
{\bf a}^R_i=c^R_i(1+b^R_i)e^{i(\delta^R_i+\phi^R_i)}, \quad \bar{\textbf{a}}^R_i = \eta^R_i c^R_i(1-b^R_i)e^{i(\delta^R_i-\phi^R_i)},
\end{equation}
where $c^R_i$, $b^R_i$, $\delta^R_i$ and $\phi^R_i$ are real numbers that can be determined in the experiment. Note that $b^R_i$ and $\phi^R_i$ are
\CP violating, while $c^R_i$ and $\delta^R_i$ are \CP conserving.
The direct \CP asymmetry for a particular intermediate state $R$ with the transversity $i$ component is
\begin{equation}
A_{\CP}(R)_i \equiv \frac{|\bar{\textbf{a}}^R_i|^2-|{\bf a}^R_i|^2}{|\bar{\textbf{a}}^R_i|^2+|{\bf a}^R_i|^2}=\frac{-2b_i^R}{1+(b^{R}_i)^2}.
\end{equation}
In the case that direct \CP violation is present, the experiment measures an ``effective" phase that is the sum of the \CP violation due to the interference between mixing the decay and direct \CP violation, given by
\begin{equation}
\phi_s^{\rm eff}(R)_i=\phi_s + 2\phi_i^R, {~\rm or~} 2\beta^{\rm eff}(R)_i = 2\beta + 2\phi_i^R.
\end{equation}

To implement this procedure data need to be fit with the probability density functions (PDFs) given in Eqs.~(\ref{eq:pdf1}) and (\ref{eq:pdf2}). The normalization can be computed by first integrating over $t$, $\angmu$ and $\chi$ analytically, then by using numerical integration for the remaining variables; the terms containing variable $\chi$ in Eqs.~(\ref{Eq:A}),~(\ref{Eq:Ab}) and~(\ref{Eq:AAb}) are zero when integrating over $\chi\in [-\pi,\pi]$. The data can be either flavour tagged or not \cite{Aaij:2012mu,*Abazov:2006qp}. In the latter case the two PDFs are averaged.

Without considering $m_{hh}$ dependence, time-dependent angular analysis for $\phi_s$ determination in $\Bs\to \jpsi \phi$ decay~\cite{LHCb-CONF-2012-002,LHCb:2011aa,CDF:2011af,*Abazov:2011ry,*:2012fu} cannot distinguish between two ambiguous solutions, one that is 
($\phi_s$, $\Delta \Gamma)$ and the other being $(\pi-\phi_s$,$-\Delta \Gamma$), because the time-dependent differential decay rates are invariant under this  transformation together with a similar transformation for the strong phases. This ambiguity has been resolved by the LHCb collaboration~\cite{Aaij:2012eq} using the P-wave $\phi$ interference with the $K^+K^-$ S-wave~\cite{Stone:2008ak} as a function of dikaon invariant mass as suggested in \cite{Xie:2009fs}. Our Dalitz-plot formalism automatically takes the strong phases as a function of $m_{hh}$ into account in the complex function $A_R(m_{hh})$ in Eq.~(\ref{eq:heart}), and thus provides only one solution for ($\phi_s$, $\Gamma_s$) without any ambiguity.

\section{Conclusions}
We have presented a method that can be used to extract the \CP violating phase for neutral $B$ meson decays into a spin-1 resonance that decays to a dilepton pair and a $\pi^+\pi^-$ or $K^+K^-$ pair, using the full set of mass and angular variables. Thus \CP violation can be measured using a much larger set of final states. For example, the $K^+K^-$ mass range in $\Bsb\to \jpsi K^+K^-$ can be used including higher mass states such as the $f_2^{\prime}(1525)$ \cite{Aaij:2011ac,*Abazov:2012dz}.

\section*{Acknowledgments}
We thank Peter Clarke, Greig Cowan, Jeroen van Leerdam, Stephanie Hansmann-Menzemer and Yuehong Xie for useful discussions. We are grateful for the support we have received from the U. S. National Science Foundation.

\section*{Appendix: Application to S- and P-waves in $\Bs \to \jpsi K^+ K^-$}
Time-dependent angular analysis~\cite{LHCb:2011aa,CDF:2011af,*Abazov:2011ry,*:2012fu} has been applied to $\Bs\to \jpsi \phi$ considering both a P-wave resonance, $\phi\to K^+K^-$, and S-wave contamination~\cite{Stone:2008ak}. Here we show that our formulation reduces to
previously used expressions \cite{Xie:2009fs,Azfar:2010nz} by considering only the $\phi$ mass region in $\Bsb\to\jpsi K^+K^-$ decays.
With this simplification, it is necessary to consider only S- and P-waves in the $K^+K^-$ system.
 Eq.~(\ref{H}) can be rewritten as
\begin{eqnarray}
\CH_0& = & \frac{H_S}{\sqrt{3}}d_{0,0}^0(\angpi)+H_0d_{0,0}^1(\angpi) = \frac{H_S}{\sqrt{3}}+H_0\cos\angpi,\nonumber\\
\CH_+ &=& H_+ d_{-1,0}^1(\angpi) = H_+\frac{\sin\angpi}{\sqrt{2}},\\
\CH_- &=& H_- d_{1,0}^1(\angpi) =-H_-\frac{\sin\angpi}{\sqrt{2}},\nonumber
\end{eqnarray}
where $H_S$ is the helicity amplitude for S-wave, and $H_{0,\pm}$ are the helicity amplitudes for P-wave  ($\lambda=0,\pm1$). Then Eq.~(\ref{Eq:A}) can be expressed as
\begin{equation}\label{Eq:Ahel}
|A_f|^2=\displaystyle\sum_{k=1}^{10} p_k G_k(\bf \Omega_{\rm hel})
\end{equation}
in terms of the amplitudes $H$, where ${\bf \Omega}_{\rm hel}$ is short hand for the three angular variables $(\angpi,\angmu,\chi)$. The individual terms for $p_k$ and $G_k(\bf \Omega_{\rm hel})$ for $k$=1--10 are listed in Table~\ref{Tab:A2}.

\begin{table}[hbtp]
\centering
\caption{Definition of the functions $p_k$ and $G_k(\bf \Omega_{\rm hel})$ of Eq.~(\ref{Eq:Ahel}). }\label{Tab:A2}
\def\arraystretch{1.2}
\begin{tabular}{|c|l|r|}\hline
$k$ & ~~~~~~~~~~ $p_k$& $G_k(\bf \Omega_{\rm hel})$~~~~~~~~~~\\\hline
1&$|\frac{H_S}{\sqrt{3}}|^2$ & $\sin^2\tone$\\
2&$|H_0|^2$& $\sin^2\tone\cos^2 \ttwo$ \\
3&$|H_+|^2+|H_-|^2$ & $\frac{1}{4}(1+\cos^2\tone)\sin^2\ttwo$\\
4&$\Real(\frac{H_S}{\sqrt{3}}H_0^*)$& $2\sin^2\tone\cos\ttwo$\\
5&$\Real(H_+H_-^*)$&$-\frac{1}{2}\sin^2\tone \sin^2\ttwo \cos2\chi$\\
6&$\Imag(H_+H_-^*)$ & $\frac{1}{2}\sin^2\tone \sin^2\ttwo \sin2\chi$\\
7&$\Real[\frac{H_S}{\sqrt{3}}(H_+^*+H_-^*)]$ & $\frac{1}{2}\sin2\tone\sin\ttwo\cos\chi$\\
8&$\Imag(\frac{H_S}{\sqrt{3}}(H_+^*-H_-^*))$& $\frac{1}{2}\sin2\tone\sin\ttwo\sin\chi$\\
9&$\Real(H_0(H_+^*+H_-^*))$& $\frac{1}{4}\sin2\tone\sin2\ttwo\cos\chi$\\
10&$\Imag(H_0(H_+^*-H_-^*))$& $-\frac{1}{4}\sin2\tone\sin2\ttwo\sin\chi$\\\hline
\end{tabular}
\end{table}

The functions can be expressed using transversity amplitudes by
using the relations between helicity and transversity amplitudes (A)~\cite{Abe:2001ks}, and the relations between helicity $(\angpi,\angmu,\chi)$, ${\bf \Omega}_{\rm hel}$, and transversity angles $(\trpsi,\trt,\trphi)$, ${\bf \Omega}_{\rm tr}$.
The amplitudes relations are
\begin{eqnarray}
A_S &=& H_S, \nonumber\\ 
A_0 &=& H_0,\nonumber\\
\Apl &=& \frac{1}{\sqrt{2}}(H_++H_-),\nonumber\\
\App &=& \frac{1}{\sqrt{2}}(H_+-H_-),
\end{eqnarray}
and the angular relationships are
\begin{eqnarray}
\cos\trpsi&=&\cos\angpi, \nonumber\\
\sin\trt\cos\trphi&=&-\cos\angmu, \nonumber\\
\sin\trt\sin\trphi&=&-\sin\angmu\cos\chi, \nonumber\\\
\cos\trt&=&\sin\angmu\sin\chi.
\end{eqnarray}

The amplitudes $\Ab_i$ are related to $A_i$  as
\begin{equation}\label{CPA}
\frac{q}{p}\frac{\Ab_i}{A_i}=\eta_i e^{-i\phi_s},
\end{equation}
where $\eta_i$ is \CP eigenvalue of the $i$ component; $\eta_S$ and $\eta_\perp=-1$, and $\eta_0$ and $\eta_\parallel=1$.
We express the amplitutes as functions of either the helicity distributions or transversity distributions as the sums 
\begin{equation}\label{Eq:Atr}
|\brabar{A}_f|^2=\displaystyle\sum_{k=1}^{10} \brabar{q}_k g_k({\bf \Omega}_{\rm hel}) = \sum_{k=1}^{10} \brabar{q}_k g_k({\bf \Omega}_{\rm tr}).
\end{equation}
From Eqs.~(\ref{Eq:AAb}) and~(\ref{CPA}), we compute the interference terms as 
\begin{equation}\label{Eq:Int}
\frac{q}{p}A_f^*\Ab_f= e^{-i\phi_s}\left(\sum_{k=1}^{10} r_k g_k(\bf \Omega_{\rm tr})\right).
\end{equation}
Each term is listed in Table~\ref{Tab:A2new}. 
\begin{table}[hbtp]
\centering
\caption{Definition of the functions used in Eqs.~(\ref{Eq:Atr}) and (\ref{Eq:Int}). When two signs appear, the upper one corresponds to $q_k$ and the lower to $\bar{q}_k$.}\label{Tab:A2new}
\def\arraystretch{1.2}
\begin{tabular}{|c|c|r|r|}\hline
$\brabar{q}_k$ & $r_k$& $g_k(\bf \Omega_{\rm hel})$~~~~~~~~~~~~& $g_k(\bf \Omega_{\rm tr})$~~~~~~~~~~~~~   \\
\hline
$|A_0|^2$& $|A_0|^2$& $\sin^2\tone\cos^2 \ttwo$ &$\cos\trpsi(1-\sin^2\trt\cos^2\trphi)$ \\
$|\Apl|^2$& $|\Apl|^2$& $\frac{1}{2}(1-\sin^2\tone\cos^2\chi)\sin^2\ttwo$&$\frac{1}{2}\sin^2\trpsi(1-\sin^2\trt\sin^2\trphi)$\\
$|\App|^2$& $-|\App|^2$&$\frac{1}{2}(1-\sin^2\tone\sin^2\chi)\sin^2\ttwo$ & $\frac{1}{2}\sin^2\trpsi\sin^2\trt$ \\
$|\frac{A_S}{\sqrt{3}}|^2$ & $-|\frac{A_S}{\sqrt{3}}|^2$ &$\sin^2\tone$ & $1-\sin^2\trt\cos^2\trphi$\\
$\Real(A_0^*\Apl)$&$\Real(A_0^*\Apl)$&$\frac{1}{2\sqrt{2}}\sin2\tone\sin2\ttwo\cos\chi$&$\frac{1}{2\sqrt{2}}\sin2\trpsi\sin^2\trt\sin2\trphi$\\
$\pm\Imag(A_0^*\App)$&$i\Real(A_0^*\App)$&$-\frac{1}{2\sqrt{2}}\sin2\tone\sin2\ttwo\sin\chi$&$\frac{1}{2\sqrt{2}}\sin2\trpsi\sin2\trt\cos\trphi$\\
$\pm\Real(A_0^*\frac{A_S}{\sqrt{3}})$&$-i\Imag(A_0^*\frac{A_S}{\sqrt{3}})$& $2\sin^2\tone\cos\ttwo$&$2\cos\trpsi(1-\sin^2\trt\cos^2\trphi)$\\
$\pm\Imag(\Apl^*\App)$&$i\Real(\Apl^*\App)$&$\frac{1}{2}\sin^2\tone\sin^2\ttwo\sin2\chi$&$-\frac{1}{2}\sin\trpsi\sin2\trt\sin\trphi$\\
$\pm\Real(\Apl^*\frac{A_S}{\sqrt{3}})$&$-i\Imag(\Apl^*\frac{A_S}{\sqrt{3}})$&$\frac{1}{\sqrt{2}}\sin2\tone\sin\ttwo\cos\chi$&$\frac{1}{\sqrt{2}}\sin\trpsi\sin^2\trt\sin2\trphi$\\
$\Imag(\App^*\frac{A_S}{\sqrt{3}})$&$-\Imag(\App^*\frac{A_S}{\sqrt{3}})$&$\frac{1}{\sqrt{2}}\sin2\tone\sin\ttwo\sin\chi$&$-\frac{1}{\sqrt{2}}\sin\trpsi\sin2\trt\cos\trphi$\\
\hline
\end{tabular}
\end{table}
In Ref.~\cite{LHCb:2011aa} the time-dependent and angular-dependent rate for $\Bs \to \jpsi \phi$ is written as
\begin{equation}
\frac{\dv^4\Gamma}{\dv t\dv {\bf \Omega_{\rm tr}}} \propto \sum_{k=1}^{10} h_k(t)f_k({\bf \Omega_{\rm tr}}),
\end{equation}
where the time-dependent function
\begin{equation}
h_k(t)=N_k e^{-\Gamma t}[a_k \ch + c_k \cs + b_k\sh + d_k\sn],
\end{equation}
and the $f_k({\bf \Omega_{\rm tr}})$ represent angular-dependent functions.
Comparing with Eq.~(\ref{Eq-t}) it can be seen that $a_k$ corresponds to $\displaystyle \frac{|A_f|^2+|\Ab_f|^2}{2}$, $c_k$ to $\displaystyle \frac{|A_f|^2-|\Ab_f|^2}{2}$,
$b_k$ to $-\Real(\frac{q}{p}A_f^*\Ab_f)$ and $d_k$ to  $-\Imag(\frac{q}{p}A_f^*\Ab_f)$. Using Table~\ref{Tab:A2new}, we find the same equations as shown in Ref.~\cite{LHCb:2011aa}.



\begin{mcitethebibliography}{10}
\mciteSetBstSublistMode{n}
\mciteSetBstMaxWidthForm{subitem}{\alph{mcitesubitemcount})}
\mciteSetBstSublistLabelBeginEnd{\mcitemaxwidthsubitemform\space}
{\relax}{\relax}

\bibitem{PDG}
Particle Data Group, J.~Beringer {\em et~al.},
  \ifthenelse{\boolean{articletitles}}{{\it {Review of Particle Physics
  (RPP)}}, }{}\href{http://dx.doi.org/10.1103/PhysRevD.86.010001}{Phys.\ Rev.\
  {\bf D86} (2012) 010001}\relax
\mciteBstWouldAddEndPuncttrue
\mciteSetBstMidEndSepPunct{\mcitedefaultmidpunct}
{\mcitedefaultendpunct}{\mcitedefaultseppunct}\relax
\EndOfBibitem
\bibitem{LHCb-CONF-2012-002}
LHCb collaboration, R.~Aaij {\em et~al.},
  \ifthenelse{\boolean{articletitles}}{{\it Tagged time-dependent angular
  analysis of \decay{\Bs}{\jpsi\phi} decays at \lhcb}, }{}
  \href{http://cdsweb.cern.ch/search?p=LHCb-CONF-2012-002&f=reportnumber&actio%
n_search=Search&c=LHCb+Reports&c=LHCb+Conference+Proceedings&c=LHCb+Conference%
+Contributions&c=LHCb+Notes&c=LHCb+Theses&c=LHCb+Papers}
  {LHCb-CONF-2012-002}\relax
\mciteBstWouldAddEndPuncttrue
\mciteSetBstMidEndSepPunct{\mcitedefaultmidpunct}
{\mcitedefaultendpunct}{\mcitedefaultseppunct}\relax
\EndOfBibitem
\bibitem{LHCb:2011aa}
LHCb collaboration, R.~Aaij {\em et~al.},
  \ifthenelse{\boolean{articletitles}}{{\it {Measurement of the CP-violating
  phase $\phi_s$ in the decay $\Bs \to J/\psi \phi$}},
  }{}\href{http://dx.doi.org/10.1103/PhysRevLett..108.101803}{Phys.\ Rev.\
  Lett.\  {\bf 108} (2012) 101803}, \href{http://arxiv.org/abs/1112.3183}{{\tt
  arXiv:1112.3183}}\relax
\mciteBstWouldAddEndPuncttrue
\mciteSetBstMidEndSepPunct{\mcitedefaultmidpunct}
{\mcitedefaultendpunct}{\mcitedefaultseppunct}\relax
\EndOfBibitem
\bibitem{CDF:2011af}
CDF collaboration, T.~Aaltonen {\em et~al.},
  \ifthenelse{\boolean{articletitles}}{{\it {Measurement of the CP-violating
  phase $\beta_s^{J/\Psi\phi}$ in $B^0_s \to J/\Psi \phi$ decays with the CDF
  II Detector}}, }{}\href{http://dx.doi.org/10.1103/PhysRevD.85.072002}{Phys.\
  Rev.\  {\bf D85} (2012) 072002}, \href{http://arxiv.org/abs/1112.1726}{{\tt
  arXiv:1112.1726}}\relax
\mciteBstWouldAddEndPuncttrue
\mciteSetBstMidEndSepPunct{\mcitedefaultmidpunct}
{\mcitedefaultendpunct}{\mcitedefaultseppunct}\relax
\EndOfBibitem
\bibitem{Abazov:2011ry}
D0 collaboration, V.~M. Abazov {\em et~al.},
  \ifthenelse{\boolean{articletitles}}{{\it {Measurement of the CP-violating
  phase $\phi_s^{J/\psi \phi}$ using the flavor-tagged decay $B_s^0 \rightarrow
  J/\psi \phi$ in 8 fb$^{-1}$ of $p \overline p$ collisions}},
  }{}\href{http://dx.doi.org/10.1103/PhysRevD.85.032006}{Phys.\ Rev.\  {\bf
  D85} (2012) 032006}, \href{http://arxiv.org/abs/1109.3166}{{\tt
  arXiv:1109.3166}}\relax
\mciteBstWouldAddEndPuncttrue
\mciteSetBstMidEndSepPunct{\mcitedefaultmidpunct}
{\mcitedefaultendpunct}{\mcitedefaultseppunct}\relax
\EndOfBibitem
\bibitem{:2012fu}
ATLAS collaboration, G.~Aad {\em et~al.},
  \ifthenelse{\boolean{articletitles}}{{\it {Time-dependent angular analysis of
  the decay $\Bs \to J/\psi \phi$ and extraction of $\Delta \Gamma_s$ and the
  $CP$-violating weak phase $\phi_s$ by ATLAS}},
  }{}\href{http://arxiv.org/abs/1208.0572}{{\tt arXiv:1208.0572}}\relax
\mciteBstWouldAddEndPuncttrue
\mciteSetBstMidEndSepPunct{\mcitedefaultmidpunct}
{\mcitedefaultendpunct}{\mcitedefaultseppunct}\relax
\EndOfBibitem
\bibitem{LHCb:2012ad}
LHCb collaboration, R.~Aaij {\em et~al.},
  \ifthenelse{\boolean{articletitles}}{{\it {Measurement of the CP-violating
  phase $\phi_s$ in $\Bsb \to J/\psi \pi^+\pi^-$ decays}},
  }{}\href{http://dx.doi.org/10.1016/j.physletb.2012.06.032}{Phys.\ Lett.\
  {\bf B713} (2012) 378}, \href{http://arxiv.org/abs/1204.5675}{{\tt
  arXiv:1204.5675}}\relax
\mciteBstWouldAddEndPuncttrue
\mciteSetBstMidEndSepPunct{\mcitedefaultmidpunct}
{\mcitedefaultendpunct}{\mcitedefaultseppunct}\relax
\EndOfBibitem
\bibitem{LHCb:2011ab}
LHCb collaboration, R.~Aaij {\em et~al.},
  \ifthenelse{\boolean{articletitles}}{{\it {Measurement of the \CP violating
  phase $\phi_s$ in $\Bsb\to J/\psi f_0(980)$}},
  }{}\href{http://dx.doi.org/10.1016/j.physletb.2012.01.017}{Phys.\ Lett.\
  {\bf B707} (2012) 497}, \href{http://arxiv.org/abs/1112.3056}{{\tt
  arXiv:1112.3056}}\relax
\mciteBstWouldAddEndPuncttrue
\mciteSetBstMidEndSepPunct{\mcitedefaultmidpunct}
{\mcitedefaultendpunct}{\mcitedefaultseppunct}\relax
\EndOfBibitem
\bibitem{LHCb:2012ae}
LHCb collaboration, R.~Aaij {\em et~al.},
  \ifthenelse{\boolean{articletitles}}{{\it {Analysis of the resonant
  components in $\Bsb\to\jpsi\pi^+\pi^-$ }},
  }{}\href{http://dx.doi.org/10.1103/PhysRevD.86.052006}{Phys.\ Rev.\  {\bf
  D86} (2012) 052006}, \href{http://arxiv.org/abs/1204.5643}{{\tt
  arXiv:1204.5643}}\relax
\mciteBstWouldAddEndPuncttrue
\mciteSetBstMidEndSepPunct{\mcitedefaultmidpunct}
{\mcitedefaultendpunct}{\mcitedefaultseppunct}\relax
\EndOfBibitem
\bibitem{Dighe:1995pd}
A.~S. Dighe, I.~Dunietz, H.~J. Lipkin, and J.~L. Rosner,
  \ifthenelse{\boolean{articletitles}}{{\it {Angular distributions and lifetime
  differences in $B_s \to J/\psi \phi$ decays}},
  }{}\href{http://dx.doi.org/10.1016/0370-2693(95)01523-X}{Phys.\ Lett.\  {\bf
  B369} (1996) 144}, \href{http://arxiv.org/abs/hep-ph/9511363}{{\tt
  arXiv:hep-ph/9511363}}\relax
\mciteBstWouldAddEndPuncttrue
\mciteSetBstMidEndSepPunct{\mcitedefaultmidpunct}
{\mcitedefaultendpunct}{\mcitedefaultseppunct}\relax
\EndOfBibitem
\bibitem{Dighe:1998vk}
A.~S. Dighe, I.~Dunietz, and R.~Fleischer,
  \ifthenelse{\boolean{articletitles}}{{\it {Extracting CKM phases and $B_s -
  \bar{B}_s$ mixing parameters from angular distributions of nonleptonic $B$
  decays}}, }{}\href{http://dx.doi.org/10.1007/s100520050372}{Eur.\ Phys.\ J.\
  {\bf C6} (1999) 647}, \href{http://arxiv.org/abs/hep-ph/9804253}{{\tt
  arXiv:hep-ph/9804253}}\relax
\mciteBstWouldAddEndPuncttrue
\mciteSetBstMidEndSepPunct{\mcitedefaultmidpunct}
{\mcitedefaultendpunct}{\mcitedefaultseppunct}\relax
\EndOfBibitem
\bibitem{Dalitz:1953cp}
R.~Dalitz, \ifthenelse{\boolean{articletitles}}{{\it {On the analysis of
  $\tau$-meson data and the nature of the $\tau$-meson}},
  }{}\href{http://dx.doi.org/10.1080/14786441008520365}{Phil.\ Mag.\  {\bf 44}
  (1953) 1068}\relax
\mciteBstWouldAddEndPuncttrue
\mciteSetBstMidEndSepPunct{\mcitedefaultmidpunct}
{\mcitedefaultendpunct}{\mcitedefaultseppunct}\relax
\EndOfBibitem
\bibitem{Nierste:2009wg}
U.~Nierste, \ifthenelse{\boolean{articletitles}}{{\it {Three lectures on Meson
  mixing and CKM phenomenology}}, }{}\href{http://arxiv.org/abs/0904.1869}{{\tt
  arXiv:0904.1869}}\relax
\mciteBstWouldAddEndPuncttrue
\mciteSetBstMidEndSepPunct{\mcitedefaultmidpunct}
{\mcitedefaultendpunct}{\mcitedefaultseppunct}\relax
\EndOfBibitem
\bibitem{Bigi:2000yz}
I.~I. Bigi and A.~Sanda, \ifthenelse{\boolean{articletitles}}{{\it {CP
  violation}}, }{}Camb.\ Monogr.\ Part.\ Phys.\ Nucl.\ Phys.\ Cosmol.\  {\bf 9}
  (2000) 1\relax
\mciteBstWouldAddEndPuncttrue
\mciteSetBstMidEndSepPunct{\mcitedefaultmidpunct}
{\mcitedefaultendpunct}{\mcitedefaultseppunct}\relax
\EndOfBibitem
\bibitem{Dunietz:1990cj}
I.~Dunietz {\em et~al.}, \ifthenelse{\boolean{articletitles}}{{\it {How to
  extract CP violating asymmetries from angular correlations}},
  }{}\href{http://dx.doi.org/10.1103/PhysRevD.43.2193}{Phys.\ Rev.\  {\bf D43}
  (1991) 2193}\relax
\mciteBstWouldAddEndPuncttrue
\mciteSetBstMidEndSepPunct{\mcitedefaultmidpunct}
{\mcitedefaultendpunct}{\mcitedefaultseppunct}\relax
\EndOfBibitem
\bibitem{Mizuk:2008me}
Belle collaboration, R.~Mizuk {\em et~al.},
  \ifthenelse{\boolean{articletitles}}{{\it {Observation of two resonance-like
  structures in the $\pi^+ \chi_{c1}$ mass distribution in exclusive
  $\overline{B}^0\to K^- \pi^+ \chi_{c1}$ decays}},
  }{}\href{http://dx.doi.org/10.1103/PhysRevD.78.072004}{Phys.\ Rev.\  {\bf
  D78} (2008) 072004}, \href{http://arxiv.org/abs/0806.4098}{{\tt
  arXiv:0806.4098}}\relax
\mciteBstWouldAddEndPuncttrue
\mciteSetBstMidEndSepPunct{\mcitedefaultmidpunct}
{\mcitedefaultendpunct}{\mcitedefaultseppunct}\relax
\EndOfBibitem
\bibitem{Aaij:2012mu}
LHCb Collaboration, R.~Aaij {\em et~al.},
  \ifthenelse{\boolean{articletitles}}{{\it {Opposite-side flavour tagging of
  $B$ mesons at the LHCb experiment}},
  }{}\href{http://dx.doi.org/10.1140/epjc/s10052-012-2022-1}{Eur.\ Phys.\ J.\
  {\bf C72} (2012) 2022}, \href{http://arxiv.org/abs/1202.4979}{{\tt
  arXiv:1202.4979}}\relax
\mciteBstWouldAddEndPuncttrue
\mciteSetBstMidEndSepPunct{\mcitedefaultmidpunct}
{\mcitedefaultendpunct}{\mcitedefaultseppunct}\relax
\EndOfBibitem
\bibitem{Abazov:2006qp}
D0 Collaboration, V.~Abazov {\em et~al.},
  \ifthenelse{\boolean{articletitles}}{{\it {Measurement of $B_d$ mixing using
  opposite-side flavor tagging}},
  }{}\href{http://dx.doi.org/10.1103/PhysRevD.74.112002}{Phys.\ Rev.\  {\bf
  D74} (2006) 112002}, \href{http://arxiv.org/abs/hep-ex/0609034}{{\tt
  arXiv:hep-ex/0609034}}\relax
\mciteBstWouldAddEndPuncttrue
\mciteSetBstMidEndSepPunct{\mcitedefaultmidpunct}
{\mcitedefaultendpunct}{\mcitedefaultseppunct}\relax
\EndOfBibitem
\bibitem{Aaij:2012eq}
LHCb Collaboration, R.~Aaij {\em et~al.},
  \ifthenelse{\boolean{articletitles}}{{\it {Determination of the sign of the
  decay width difference in the $B_s$ system}},
  }{}\href{http://dx.doi.org/10.1103/PhysRevLett.108.241801}{Phys.\ Rev.\
  Lett.\  {\bf 108} (2012) 241801}, \href{http://arxiv.org/abs/1202.4717}{{\tt
  arXiv:1202.4717}}\relax
\mciteBstWouldAddEndPuncttrue
\mciteSetBstMidEndSepPunct{\mcitedefaultmidpunct}
{\mcitedefaultendpunct}{\mcitedefaultseppunct}\relax
\EndOfBibitem
\bibitem{Stone:2008ak}
S.~Stone and L.~Zhang, \ifthenelse{\boolean{articletitles}}{{\it {S-waves and
  the measurement of CP violating phases in $B_s$ decays}},
  }{}\href{http://dx.doi.org/10.1103/PhysRevD.79.074024}{Phys.\ Rev.\  {\bf
  D79} (2009) 074024}, \href{http://arxiv.org/abs/0812.2832}{{\tt
  arXiv:0812.2832}}\relax
\mciteBstWouldAddEndPuncttrue
\mciteSetBstMidEndSepPunct{\mcitedefaultmidpunct}
{\mcitedefaultendpunct}{\mcitedefaultseppunct}\relax
\EndOfBibitem
\bibitem{Xie:2009fs}
Y.~Xie, P.~Clarke, G.~Cowan, and F.~Muheim,
  \ifthenelse{\boolean{articletitles}}{{\it {Determination of 2$\beta_s$ in
  $\Bs\to \jpsi K^+ K^-$ Decays in the Presence of a $K^+ K^-$ S-Wave
  Contribution}},
  }{}\href{http://dx.doi.org/10.1088/1126-6708/2009/09/074}{JHEP {\bf 0909}
  (2009) 074}, \href{http://arxiv.org/abs/0908.3627}{{\tt
  arXiv:0908.3627}}\relax
\mciteBstWouldAddEndPuncttrue
\mciteSetBstMidEndSepPunct{\mcitedefaultmidpunct}
{\mcitedefaultendpunct}{\mcitedefaultseppunct}\relax
\EndOfBibitem
\bibitem{Aaij:2011ac}
LHCb Collaboration, R.~Aaij {\em et~al.},
  \ifthenelse{\boolean{articletitles}}{{\it {Observation of $B_s \to J/\psi
  f^\prime_2(1525)$ in $J/\psi K^+K^-$ final states}},
  }{}\href{http://dx.doi.org/10.1103/PhysRevLett.108.151801}{Phys.\ Rev.\
  Lett.\  {\bf 108} (2012) 151801}, \href{http://arxiv.org/abs/1112.4695}{{\tt
  arXiv:1112.4695}}\relax
\mciteBstWouldAddEndPuncttrue
\mciteSetBstMidEndSepPunct{\mcitedefaultmidpunct}
{\mcitedefaultendpunct}{\mcitedefaultseppunct}\relax
\EndOfBibitem
\bibitem{Abazov:2012dz}
D0 Collaboration, V.~M. Abazov {\em et~al.},
  \ifthenelse{\boolean{articletitles}}{{\it {Study of the decay $B_s^0
  \rightarrow J/\psi f_2^{\prime}(1525)$ in $\mu^+ \mu^- K^+K^-$ final
  states}}, }{}Phys.\ Rev.\ D (2012) \href{http://arxiv.org/abs/1204.5723}{{\tt
  arXiv:1204.5723}}\relax
\mciteBstWouldAddEndPuncttrue
\mciteSetBstMidEndSepPunct{\mcitedefaultmidpunct}
{\mcitedefaultendpunct}{\mcitedefaultseppunct}\relax
\EndOfBibitem
\bibitem{Azfar:2010nz}
F.~Azfar {\em et~al.}, \ifthenelse{\boolean{articletitles}}{{\it {Formulae for
  the Analysis of the Flavor-Tagged Decay $\Bs\to \jpsi\phi$}},
  }{}\href{http://dx.doi.org/10.1007/JHEP11(2010)158}{JHEP {\bf 1011} (2010)
  158}, \href{http://arxiv.org/abs/1008.4283}{{\tt arXiv:1008.4283}}\relax
\mciteBstWouldAddEndPuncttrue
\mciteSetBstMidEndSepPunct{\mcitedefaultmidpunct}
{\mcitedefaultendpunct}{\mcitedefaultseppunct}\relax
\EndOfBibitem
\bibitem{Abe:2001ks}
K.~Abe, M.~Satpathy, and H.~Yamamoto, \ifthenelse{\boolean{articletitles}}{{\it
  {Time dependent angular analyses of $B$ decays}},
  }{}\href{http://arxiv.org/abs/hep-ex/0103002}{{\tt
  arXiv:hep-ex/0103002}}\relax
\mciteBstWouldAddEndPuncttrue
\mciteSetBstMidEndSepPunct{\mcitedefaultmidpunct}
{\mcitedefaultendpunct}{\mcitedefaultseppunct}\relax
\EndOfBibitem
\end{mcitethebibliography}
\ifx\mcitethebibliography\mciteundefinedmacro
\PackageError{LHCb.bst}{mciteplus.sty has not been loaded}
{This bibstyle requires the use of the mciteplus package.}\fi
\providecommand{\href}[2]{#2}

\end{document}